\documentstyle[multicol,prl,aps,floats,epsf]{revtex} 
\begin{document} 
\draft 
\preprint{ } 
\wideabs{
\title{Isotropic dispersion, line shape, and remnant Fermi surface in one hole 
problem } 
\author{Z. Y. Weng, D. N. Sheng, and C. S. Ting} 
\address{Texas Center for Superconductivity and Department of Physics\\ 
University of Houston, Houston, TX 77204-5506 }  
\maketitle 
\date{today}
\begin{abstract} 
It is shown that the isotropic dispersion relation, line shape of the spectral function, and remnant Fermi surface found in recent photoemission 
experiment in insulating Ca$_2$CuO$_2$Cl$_2$ by Ronning {\it et al.}
(Science, {\bf 282}, 2067 (1998)) can be consistently explained by a {\it 
direct} holon hopping process which fundamentally undermines the conventional 
self-consistent Born approximation approach to the $t-J$ model. 
\end{abstract} 
 }

The problem of one hole moving in the antiferromagnetic (AF) spin background 
can give us some important clue about the finite doping. For many years, the 
self-consistent Born approximation (SCBA) method\cite{scba,kane,scba2,scba3} 
based on the $t-J$ model has 
provided a standard description of the one hole problem, consistent with exact 
numerical calculations\cite{diagonalization} up to 32 sites\cite{leung}, which 
predicts a {\it sharp} quasiparticle 
coherent peak in the spectral function with an {\it anisotropic} dispersion 
relation around the Fermi points ${\bf k}_0=(\pm \pi/2, \pm\pi/2)$.

But the SCBA results are not consistent with angle-resolved
photoemission spectroscopy (ARPES) measurements in the parent compounds of 
high-$T_c$ superconductors: Sr$_2$CuO$_2$Cl$_2$\cite{wells,t'-t''} and 
Ca$_2$CuO$_2$Cl$_2$ \cite{ronning}, which indicate that while the 
``quasiparticle'' peak is {\it not} sharp at all with an intrinsic broad
energy structure, its dispersion is fairly {\it isotropic} around the Fermi
point ${\bf k}_0$. A further challenge to theory comes from the recent
high-quality measurement\cite{ronning} revealing the so-called remnant Fermi 
surface (RFS) in the electron {\it relative} momentum distribution $n^r_k$, 
in contrast to the low-energy Fermi point structure.

To explain the isotropy of the dispersion relation, the second and third 
neighbor hopping $t'$ and $t''$ adding to the $t-J$ model were suggested in 
literature\cite{t'-t'',t',tohyama} which could improve the agreement between 
the theory and experiment under some proper choice of parameters. 
But Laughlin\cite{laughlin} has proposed a different view in which the isotropic
dispersion may come from a more intrinsic origin: the spinon spectrum. In this
spin-charge separation picture, the size of the dispersion is also 
accounted for by the superexchange $J$ without $t'$ and 
$t''$, and the injected hole decaying into a spinon-holon pair can provide a
natural broad structure in the spectral function. Such a spin-charge separation 
scenario is fundamentally different from the Landau quasiparticle picture in the 
SCBA approach.   

The purpose of this paper is to reexamine the one hole issue based on the 
$t-J$ model. We show that due to the existence of a {\it direct} holon hopping 
term missing in the SCBA scheme, the low-energy physics can be completely 
changed from the SCBA prediction, and the isotropic ``quasiparticle'' 
dispersion, line shape of the spectral function, as well as the remnant 
Fermi surface structure observed in experiment can all be explained as the 
consequence of the spin-charge separation, which in some main aspect is consistent with the proposal made in Ref.\onlinecite{laughlin}. 

We start with the slave-fermion representation $c_{i\sigma}=f^{\dagger}_ib_{i\sigma}$\cite{kane}. In the $t-J$ model 
$H_{t-J}=H_t + H_J$, $H_J=-\frac J 2 \sum_{\langle ij\rangle} 
(\hat{\Delta}^s_{ij})^{\dagger}\hat{\Delta}^s_{ij}$
with $\hat{\Delta}^s_{ij}=\sum_{\sigma}\bar{b}_{i\sigma}\bar{b}_{j-\sigma}$.
[Here $\bar{b}_{i\sigma}\equiv (-\sigma)^ib_{i\sigma}$, with $(-
\sigma)^i$ explicitly keeping track of the Marshall sign.] The 
motion of a doped hole is governed by $H_t=-t\sum_{\langle ij\rangle} 
\hat{B}_{ji} f^{\dagger}_if_j + H.c.$ with 
$\hat{B}_{ji}=\sum_{\sigma}\sigma \bar{b}^{\dagger}_{j\sigma}\bar{b}_{i\sigma}$. 
Even though the Schwinger-boson mean-field theory (SBMFT)\cite{aa} works quite 
well at half-filling, characterized by the bosonic resonating-valence-bond (RVB) 
order parameter $\langle \hat{\Delta}^s_{ij}\rangle $, the one hole problem is 
highly nontrivial. The reason can be attributed to the fact that
the hopping integral $\langle \hat{B}_{ji}\rangle=0$ in $H_t$. In
the SCBA approach, the hole has to acquire its kinetic energy through the {\it
dynamic fluctuations} of $\hat{B}_{ji}$ as follows. To the leading order in a
large-S (spin wave) expansion, one may write\cite{kane}
\begin{equation}\label{b1}
\hat{B}_{ji}\approx \hat{B}_{ji}^{LSW}= b_0\sum_{\sigma}\sigma \left[\bar{b}^{\dagger}_{j\sigma}+
\bar{b}_{i\sigma}\right],
\end{equation}
where $b_0$ denotes the condensed part of the Schwinger boson field, corresponding to AF long-range order (AFLRO) \cite{note}. Then a hole
created by $\bar{c}_{i\sigma}=f^{\dagger}_ib_0(-\sigma)^i$ can propagate by an
SCBA assisted hopping and behave just like a Landau quasiparticle
with four Fermi points at ${\bf k}_0$\cite{scba,kane,scba2,scba3}. 

However, an important process is missing in the above picture in the spin-1/2 case. We will show below that due to the RVB nature of the SBMFT description\cite{aa}, a {\em direct} hopping term for the holon actually exists 
but is omitted in (\ref{b1}). For the sake of clarity, one may set the
temperature $T$ at $0^+$ where $b_0$ (and the AFLRO) is switched off such that 
(\ref{b1}) vanishes. Then one can rewrite 
$\hat{B}_{ji}$ in the following form $\hat{B}_{ji}=\left(B^0_{ji}\right) e^{i\hat{\phi}_{ji}}$ where $B^0_{ji}=\sum_{\sigma}\bar{b}^{\dagger}_{j\sigma}{\bar b}_{i\sigma}$ and
$\hat{\phi}_{ji}= \pm(\pi/2) [1 - \sigma_{ji}]$ [$\sigma_{ji}=\pm 1 $ depending 
on $\uparrow$ or $\downarrow$ spinon exchanged with the 
holon at the given link $(ij)$]. If $\langle B^0_{ji}\rangle= B^0 
\neq 0$, the following {\it bare} holon hopping term will emerge   
\begin{equation}\label{hh}
H_h=-t_{eff}\sum_{\langle ij\rangle}  e^{
i\hat{\phi}_{ji}}{f}^{\dagger}_i{f}_j + H.c. ,
\end{equation}
with $t_{eff}\equiv tB^0$. Here $\langle \hat{B}_{ji}\rangle=0$ is entirely
due to $\langle e^{i\hat{\phi}_{ji}}\rangle=0$ but the correct measuring
of the gauge phase strength should be the gauge-invariant quantity 
$e^{i\sum_{\Gamma} \hat{\phi}_{ji}}$ for a closed path $\Gamma$. One may
write $e^{i\sum_{\Gamma} \hat{\phi}_{ji}}=(-1)^{N^{\downarrow}_{\Gamma}}$
with $N^{\downarrow}_{\Gamma}$ denoting the total number of $\downarrow$ spins
(or $\uparrow$ spins by symmetry) ``exchanged'' with the hole along the path 
$\Gamma$. Here $(-1)^{N^{\downarrow}_{\Gamma}}$ is nothing  
but the phase string factor first discussed in Ref. \onlinecite{string1} in a 
rigorous formulation.

The proof of $B^0\neq 0$ is straightforward. At half-filling, one has the Bogoliubov 
transformation
\begin{equation}\label{bogo}
\bar{b}_{i\sigma}=\frac 1 {\sqrt{N}}\sum_{\bf k}\eta_{{\bf k}\sigma} 
e^{i\sigma {\bf k}\cdot
{\bf r}_i} \left[u_{\bf k}\gamma_{{\bf k}\sigma}-v_{\bf k}\gamma_{{\bf k}-
\sigma}^{\dagger}\right]
\end{equation}
in which $\gamma_{{\bf k}\sigma}$ represents an SBMFT elementary excitation 
operator, with a gapless spectrum at $T=0^+$\cite{aa}  
\begin{equation}\label{sp}
E_k=2.32 J\sqrt{1-\xi_{\bf k}^2},
\end{equation}
and $u_{\bf k}^2-v_{\bf k}^2=1$, $2u_{\bf k}v_{\bf k}=
\xi_{\bf k}/\sqrt{1-\xi_{\bf k}^2}$ where $\xi_{\bf k}=(\cos k_x +\cos k_y)/2$.
It is crucial to recognize that the factor $\eta_{\bf k\sigma}$ in
(\ref{bogo}) which satisfies $\eta_{\sigma}=\eta_{-\sigma}^*$ and $|\eta|=1$  cannot be completely determined at the mean-field level. Specifically, the 
mean-field order parameter $\langle \bar{b}_{i\sigma}\bar{b}_{j-\sigma}\rangle$ 
is independent of different choices of $\eta_{k\sigma}$. Then one may exploit 
this symmetry for the one hole case. At each fixed hole position, there is a 
sub-Hilbert-space for spins which at the mean-field level is described by the 
Bogoliubov transformation (\ref{bogo}). By choosing $\eta_{\bf k\sigma}=
[-{\mbox sgn}(\xi_{\bf k})]^{k_h}$ where $k_h=0$ if the hole is 
on the even sublattice and $k_h =1$ if the hole is on the odd sublattice 
site, it is easy to find that ${B}^0=\frac 2 N \sum_{{\bf k}\neq 0} |\xi_{\bf k}| v_{\bf k}^2\approx 0.4$ in $H_t$ where a hole
changing sublattices is involved, while the mean-filed result remains the same 
for the superexchange term $H_J$. Note that such a hole-position-dependent 
choice of the phase $\eta_{{\bf k}\sigma}$ is perfectly legitimate 
as in the sub-Hilbert-space in which the hole position is given, the 
commutation relations of $\bar{b}$, $\bar{b}^{\dagger}$ remain unchanged. 

The existence of a bare hopping term $H_h$ in (\ref{hh}) will fundamentally 
change the SCBA result. Note that one may express $c_{i\sigma}=\bar{c}_{i\sigma}+\tilde{c}_{i\sigma}$ with $\tilde{c}_{i\sigma}=
f_i^{\dagger}:b_{i\sigma}:$ with $:b_{i\sigma}:\equiv b_{i\sigma}-b_0(-
\sigma)^i$. Due to (\ref{hh}), the propagator for $\tilde{c}_{i\sigma}$ has 
a leading contribution $\tilde{G}_{e}\approx iG_f\cdot G_b$ which reflects the 
spin-charge separation, where $G_f=-i\langle T_t (f_i^{\dagger}(t)f_j(0))\rangle$ and $G_b =-i\langle T_t (:b_{i\sigma}(t)::b^{\dagger}_{j\sigma}(0):)\rangle$. On the other hand, the 
Hamiltonian corresponding to the linear spin-wave approximation (\ref{b1}) may 
be rewritten as $H_h^{LSW}=-t \sum_{\langle ij\rangle \sigma}
\left(\bar{c}^{\dagger}_{i\sigma}\tilde{c}_{j\sigma}+ \tilde{c}_{i\sigma}^{
\dagger}\bar{c}_{j\sigma}\right)+ H.c. $, which literally shows a process for 
the quasiparticle $\bar{c}_{\sigma}$ to {\it decay} into a holon-spinon pair
represented by $\tilde{c}_{\sigma}$. When both holon and quasiparticle have no
bare kinetic energy, $H_h^{LSW}$ represents a virtual process and an SCBA 
treatment of it reproduces the well-known results\cite{scba,kane,scba2,scba3}. 
But once the holon acquires a bare kinetic energy whereas the quasiparticle
has none, $H_h^{LSW}$ simply describes the break-up process of a quasiparticle, 
where the propagator $\bar{G}_e$ for $\bar{c}_{\sigma}$ should become 
proportional to $\tilde{G}_e$ to the leading order approximation. Note that any 
higher order corrections to the quasiparticle self-energy also add to the holon 
self-energy so that the initial instability is always maintained. Therefore, the
SCBA approach is totally undermined once the spin-charge separation is present 
where the holon as a free particle has its bare kinetic energy.

The spin-charge separation reflected in $\tilde{G}_e$ leads to the convolution
law of the spectral function in the following form 
\begin{equation}\label{ae}
A^e({\bf k},\omega)=\frac 1 N \sum_{{\bf k}'\neq 0} v_{{\bf k}'}^2 {A}^h({\bf
k}'-{\bf k}, -E_{{\bf k}'}-\omega),
\end{equation}   
where the r.h.s. is obtained after using the SBMFT result, ${\mbox {Im}} G_b=-
\pi [u_{\bf k}^2\delta(\omega-E_{\bf k})-v_{\bf k}^2\delta(\omega+E_{\bf k})]$, 
for the spinon part, while the holon spectral function ${A}^h({\bf 
k},\omega)\neq 0$ at $\omega\geq 0$ which is to be determined by $H_h$ in 
(\ref{hh}). By noting that
\begin{equation}\label{phi}
e^{i\hat{\phi}_{ji}}=e^{i{\bf k}_0\cdot ({\bf r}_j-{\bf r}_i)}e^{-i\delta\hat{\phi}_{ji}}
\end{equation}
with $\delta\hat{\phi}_{ji}=\pm (\pi/2) \sigma_{ji}$ and 
$\langle \delta\hat{\phi}_{ji}\rangle =0$, one finds that 
${A}^h({\bf k},\omega)$ should be peaked at ${\bf k}={\bf k}_0$ at low energy,
and correspondingly the spinon spectrum $E_{{\bf k}+{\bf k}_0}$ will
essentially determine the low-energy ``edge'' of the electron spectral function 
in (\ref{ae}). Here we note that $\delta\hat{\phi}_{ji}$ will generally play a
role causing the incoherent motion of the holon as opposed to the {\it coherent} 
spinon with a $\delta$-function-like spectral function. As an approximation, we 
first simply treat $\delta\hat{\phi}_{ji}$ as a random flux field and determine
${A}^h({\bf k},\omega)$ numerically. We note that the overall shape of 
$A^e({\bf k},\omega)$ will {\em not} be sensitive to such an approximation as to 
be discussed below.

The typical behavior of $A^e({\bf k}, \omega)$ versus $\omega$ is shown in 
Fig. 1 at different ${\bf k}$'s along (0, 0) to ${\bf k}_0$.
The small arrows mark the energy of $E_{{\bf k}+{\bf k}_0}$ which is indeed 
correlated with the position of the ``coherent'' peak or more precisely the edge 
of $A^e({\bf k}, \omega)$. The line shape of the spectral function in Fig. 1 
reflects essentially the {\it
convolution} law of spinon and holon in a spin-charge separation context, which
is strikingly similar to those found in the ARPES measurements\cite{wells,t'-t'',ronning}. In particular, the spinon spectrum 
$E_{{\bf k}+{\bf k}_0}$ fits the ARPES data reasonably well over the whole 
Brillouin zone as shown in Fig. 2 along ${\bf k}_0$ ($\Sigma$) to (0,0) 
($\Gamma$) and (0,$\pi$) ($X$) (which are symmetric in our theory) with 
$J=0.14$$eV$. A 
very flat (only about $13\%$ change) $E_{{\bf k}+{\bf k}_0}$ along (0,0) to (0, 
$\pi$) is also in good agreement with the experiment. Therefore, the 
theory can naturally account for the ARPES without $t'$, $t''$ terms.  

In Fig. 1, the holon spectral function is computed at 
$t_{eff}= 2J$ ($t\approx 5J$) with the random (white noise) flux strength chosen 
between $\pm 0.1 \pi$ per plaquette in $48\times 48$ lattice. But we emphasize 
that the general line shape in Fig. 1 looks {\it insensitive} to those 
parameters which mainly affect $A^h$. This is because the broad feature of the 
spectral function is chiefly due to the convolution law which reflects the 
spin-charge 
separation\cite{laughlin}, and the high energy part of $A^e$ is essentially 
decided by the {\it bandwidth} ($\sim 8t_{eff}$) of the holon spectrum, {\it 
not} the sharpness of $A^h$ itself. Note that the coherent factor $v^2$ and the 
``incoherence'' of the holon caused by the gauge phase in (\ref{hh}) are crucial
for the coherent spinon spectrum to ``show up'' in $A^e({\bf k}, \omega)$. In 
the one-dimensional case, the phase string also plays a role to shape the 
one-hole spectral function as discussed in Ref.\onlinecite{suzuura}.  

Beyond the random-flux treatment, the precise form of $\delta\hat{\phi}_{ji}$ 
will be important in determining the {\it relative} 
change at different momentum of an energy-integrated $A^e$: the so-called
relative momentum distribution\cite{ronning} defined by 
$n^r_k(\omega_0)=\int^{0}_{\omega_0}\frac{d\omega}{\pi}A^e(k,\omega)$. Here
if $\omega_0$ is taken to $-\infty$, $n^r_k$ will
reduce to the momentum distribution $n_k$ which should be
featureless ($=1/2$) for the one-hole case [note 
that $\langle  c^{\dagger}_{i\sigma} c_{j\sigma}\rangle= \langle 
{b}_{i\sigma}^{\dagger}f_if_j^{\dagger}b_{j\sigma}
\rangle=\delta_{ij}\langle b_{i\sigma}^{\dagger}b_{i\sigma}\rangle$ using the
non-double-occupancy constraint]. Thus, within the framework of the $t-J$ model,
the observed RFS feature in $n^r_k$\cite{ronning} for a finite $\omega_0$ ($\sim 0.5 eV$) has nothing to do with the half-filling ground-state and must solely 
come from the {\it dynamics} of the doped hole. We have already seen that the 
characteristic ``Fermi-point'' ${\bf k}_0$ arises from the leading term of the 
gauge phase $\hat{\phi}_{ji}$ in (\ref{phi}), and we will see that beyond the 
random flux approximation $\delta\hat{\phi}_{ji}$ may be responsible for
an additional momentum structure in $n^r_k$. 

The gauge phase $\hat{\phi}$ can be more precisely described as a
nonlocal phase string effect as noted before. It may be clearly seen by
formulating $G_f$ according to Ref.\onlinecite{string1} as follows 
\begin{equation}\label{gf0} 
G_f(j,i;E)=\sum_{\{c\}\{\sigma_s\}}
\langle\hat{S}_c(\{\sigma_s\}; E) \rangle (-1)^{N_c^{\downarrow}},
\end{equation}
with $\hat{S}_c \equiv \left(\prod_{s=1}^{K}G_J(t)\left[\bar{b}^{\dagger}_{m_{s-
1}\sigma_{s}}\bar{b}_{m_s\sigma_{s}}\right]\right)G_J$, $G_J=1/(E-H_J)$,
and $N^{\downarrow}_c=\sum_s (1-\sigma_{s})/2$. In $\hat{S}_c$, the 
site $m_s$ is at a path $c$ that the holon travels, which connects sites $i$ and 
$j$: $m_0=i, m_1, ..., m_{K}=j$. In particular, one can prove\cite{string1} that
$\langle\hat{S}_c(E)\rangle\geq 0$ at low energy such that the phase string
effect introduced by $(-1)^{N_c^{\downarrow}}$ is basically {\it nonrepairable}. What we need to know is how such an oscillating phase string shapes the high
energy part of $A^e$ to determine $n^r_k$. Due to its nonrepairable nature, the
{\it singular part} of the phase string effect can be explicitly built into the 
wave function through a unitary transformation\cite{string1}, leading to 
$c_{i\sigma}=a_{i\sigma}\times e^{i\Theta_i^{string}}$.
Here $a_{i\sigma}$ refers to the {\it elementary} charge and spin 
excitations whose details are not important to
our problem below. The key thing is the phase shift $\Theta_i^{string}$ which
will then modulate the electron propagator by 
\begin{equation}
e^{-i\left[\Theta_j^{string}-\Theta_i^{string}\right]}= e^{-i 
\int_{c}\cdot \hat{\bf A}^f({\bf r})}
\end{equation}
with $
\hat{\bf A}^{f}({\bf r})=\frac 1 2 \sum_{l}[\sum_{\sigma} \sigma n^b_{l
\sigma}-1] \frac{{\bf \hat{z}}\times ({\bf r}-{\bf r}_l)}{|{\bf r}-
{\bf r}_l|^2}$ in the equal-time limit. Such a phase string factor
originated from $\hat{\phi}_{ji}$ can be similarly evaluated using the method
outlined in Ref.\onlinecite{string3}, which leads to four ``incommensurate 
peaks'' at ($\pm \pi\kappa$, 0) and (0, $\pm \pi\kappa$) in momentum space with 
$\kappa \sim 1$\cite{remark}. If this oscillating factor solely decides the 
momentum structure of $n^r_k$ 
with the rest term in the propagator mainly contributing a broadening, then one 
gets a contour plot of $n^r_k$ in Fig. 3 in terms of the superposition of 
aforementioned 
four peaks (with $\kappa=0.75$) at an arbitrary broadening (which is presumably 
controlled by the energy cutoff $\omega_0$: when $\omega_0\rightarrow -\infty$, 
the broadening in the momentum space should also go to infinity such that $n_{\bf k}=1/2$). Note that the values of $n^r_{\bf k}$ here are only for {\it 
relative} comparison. We see that the overall feature is in qualitative
agreement with the experimental data\cite{ronning} whose one quarter portion is 
also shown in the inset of Fig. 3 for comparison. Experimentally\cite{ronning}  
the RFS is defined at the sharp drop of $n_k^r$ in ${\bf k}$-space.

Finally, we point out that strictly speaking the AFLRO and RVB condensations 
only exist in the thermodynamic limit and both (\ref{b1}) and (\ref{hh}) do not 
directly apply to a finite system. In a small sample, both amplitude and phase 
fluctuations in $\hat{B}_{ji}$ are expected to be important and cannot be 
distinguished like in (\ref{hh}) where $B^0\neq 0$. And in this case a generalized SCBA scheme may be constructed to account for the leading order 
effect in the hole dynamics. This should be the reason why the small-size exact 
calculations have always identified a quasiparticle, consistent with the SCBA as 
well as power-method results\cite{scba2,scba3,lee}. However, such an SCBA 
quasiparticle is expected to eventually break up into holon-spinon composite at 
large lattice so long as there is a bosonic RVB {\it condensation}\cite{aa} 
in the thermodynamic limit to warrant (\ref{hh}).
 
In conclusion, a direct holon hopping term is identified in the one
hole problem due to the RVB spin pairing in the SBMFT framework, which is 
omitted in the SCBA approach. Consequently a drastically different 
characterization of low-energy physics emerges: The SCBA quasiparticle becomes
unstable and is replaced by a holon-spinon composite in which the holon is 
{\it incoherent} due to the phase string effect associated with the bare 
hopping, while the coherent {\it spinon} spectrum determines the {\it isotropic} 
``quasiparticle'' dispersion relation. The line shape of the spectral function
simply reflects the spin-charge decomposition, while the remnant Fermi surface
is caused by the oscillating phase string effect at high energy. These
properties well explain the recent ARPES measurements within the framework of 
the $t-J$ model.     

\acknowledgments We wish to acknowledge useful discussions with A. Ferraz, T. 
Tohyama, T.K. Lee, and especially C. Kim and F. Ronning who kindly provided
their experimental data. We thank the hospitality of Centro Internacional de
Fisica da Materia Condensada of Univ. de Brasilia and the Aspen Center for
Physics where this work was initiated and finalized.  The present work was 
supported in part by Texas ARP No.3652707, the Robert A. Welch 
foundation, and TCSUH.\\

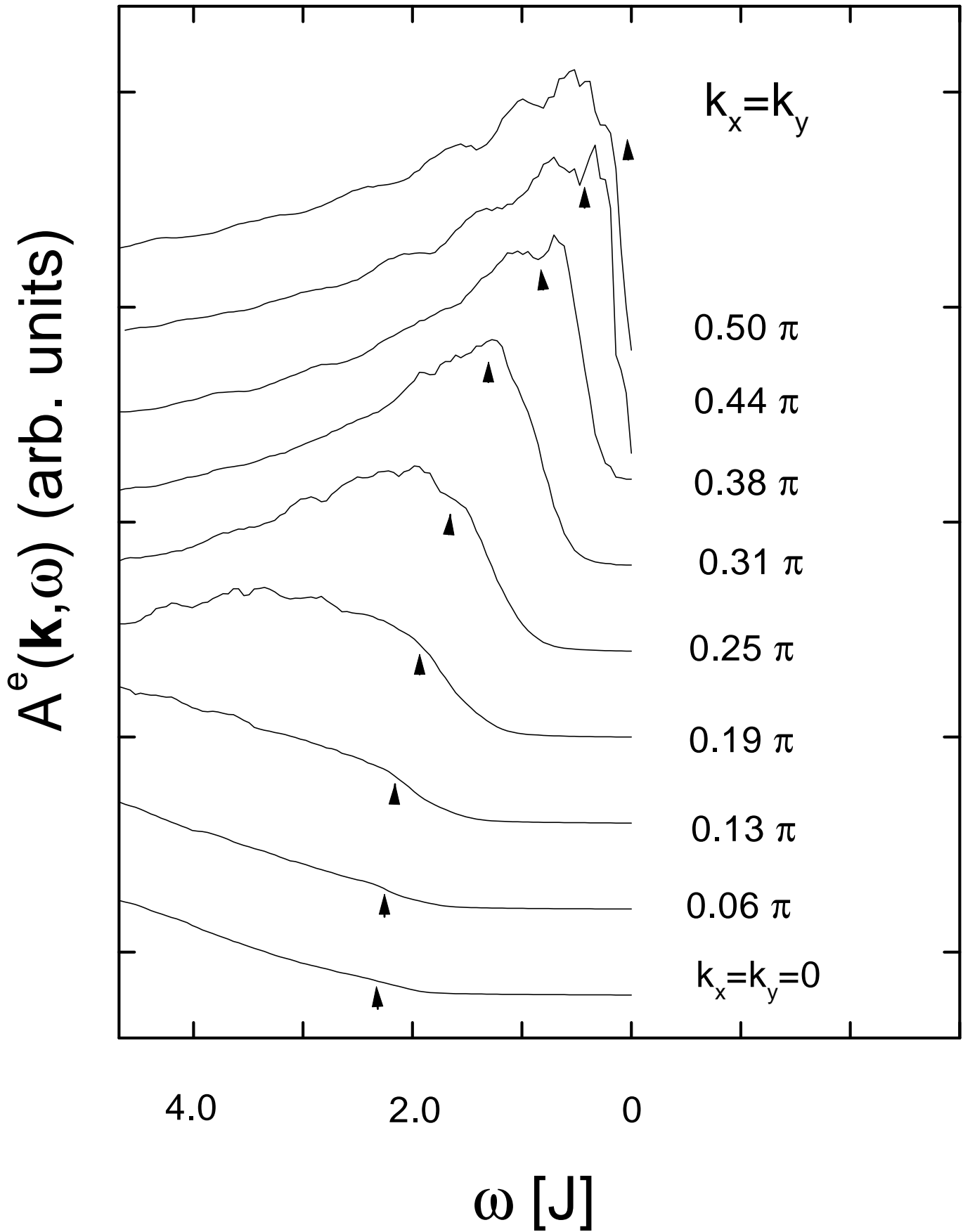
Fig. 1. The theoretical electron spectral function. The arrows mark the
corresponding energy of the isotropic spinon spectrum $E_{{\bf k}+{\bf k}_0}$.

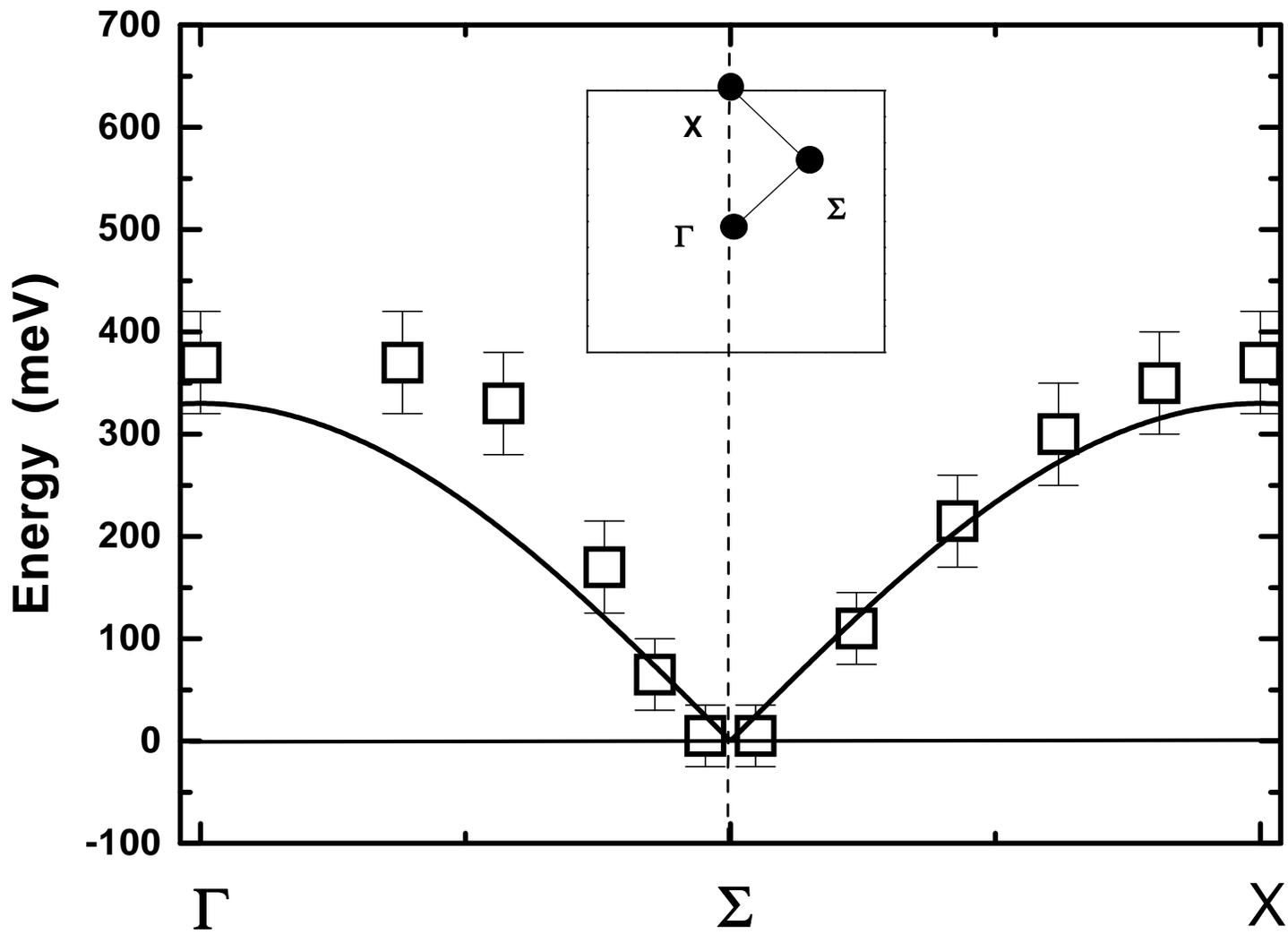
Fig. 2. The quasiparticle spectra determined by the
ARPES\cite{ronning} (open square) and by the present theory (solid curve).

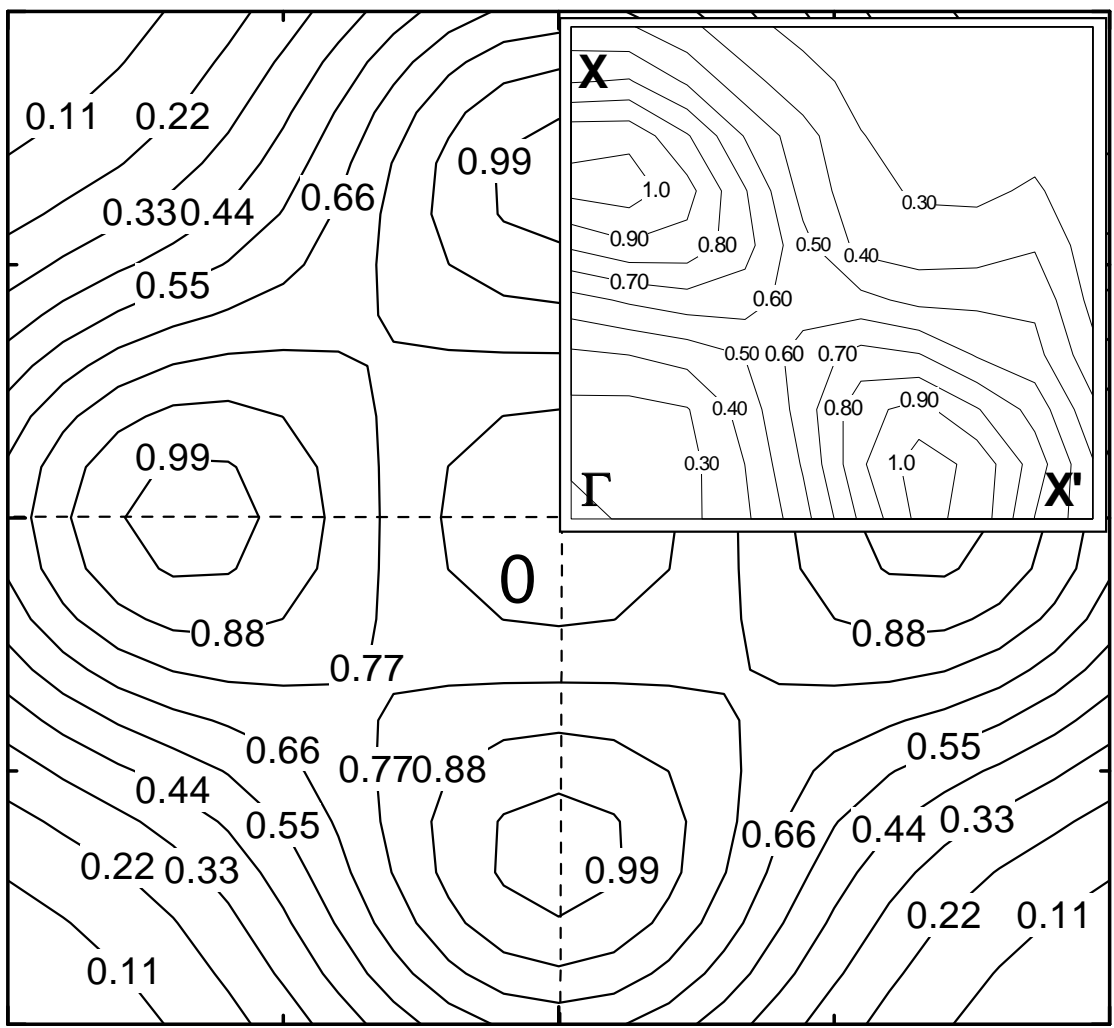
Fig. 3. The contour plot of the electron momentum distribution 
exhibiting a ``remnant Fermi surface'' structure. The experimental 
data\cite{ronning} are shown in the insert at the upper right corner.


\begin{references}
\bibitem{note} For convenience, the magnetization here is to lie in the 
$x-y$ plane instead of along the $z$-axis as in Ref.\onlinecite{kane}.
\bibitem{remark} Readers are referred to Ref.\onlinecite{string3} for the 
detail, where a similar phase-string factor appearing in the spin-spin 
correlation function is discussed which gives rise to incommensurate shifts by 
($\pm  2\pi\delta \bar{\kappa}$, 0), etc., from ($\pi$,$\pi$) with $\delta$ 
denoting the doping concentration and $\bar{\kappa}\sim 1$.
\end{references}
\end{document}